\documentstyle[12pt]{article}
\begin{document}
\title{Quantum theory for the binomial model\\in finance
theory}
\author{CHEN Zeqian\footnote{E-mail: zqchen@wipm.ac.cn}
\\{\small (Wuhan Institute of Physics and
Mathematics, CAS, P.O.Box 71010, Wuhan 430071, China)}}
\date{}
\maketitle

\begin{quote}{\bf Abstract.} In this paper, a quantum model for the
binomial market in finance is proposed. We show that its
risk-neutral world exhibits an intriguing structure as a disk in
the unit ball of ${\bf R}^3,$ whose radius is a function of the
risk-free interest rate with two thresholds which prevent
arbitrage opportunities from this quantum market. Furthermore,
from the quantum mechanical point of view we re-deduce the
Cox-Ross-Rubinstein binomial option pricing formula by considering
Maxwell-Boltzmann statistics of the system of $N$ distinguishable
particles.

\end{quote}
~\\
{\bf 1  Introduction}

Usually a problem which has two outcomes is said to be binomial.
There are many binomial problems in nature, such as the game of a
coin toss, photon's polarization and so on. It is well known that
the throw of a coin can be modelled by classical random variables,
as called Bernoulli's random variables in probability theory,
while (perhaps less well known) the mechanism of photon's
polarization however must be described by quantum mechanics. There
is a different mechanism underlying the game of a coin toss than
that of the photon's polarization, see for example Dirac [5].

In finance theory the binomial market is a useful and very popular
technique for pricing a stock option, in which only one risky
asset is binomial. Although the binomial market is very ideal
model, a realistic model may be assumed to be composed of a much
large numbers of binomial markets. This is the assumption that
underlies a widely used numerical procedure first proposed by Cox,
Ross, and Rubinstein [4], in which Bernoulli's random variables
are used to describe the only one risky asset. There seems to be
{\it a prior} no reason why the binomial market must be modelled
by using the Bernoulli's random variables, even though the
binomial market is a hypothesis and ideal model.

In this paper, a quantum model for the binomial market is
proposed. We show that its risk-neutral world exhibits an
intriguing structure as a disk in the unit ball of ${\bf R}^3,$
whose radius is a function of the risk-free interest rate with two
thresholds which prevent arbitrage opportunities from this quantum
market. Moreover, from the quantum point of view we re-deduce the
Cox-Ross-Rubinstein binomial option pricing formula by considering
Maxwell-Boltzmann statistics of the system of $N$ distinguishable
particles. Therefore, it seems that it is of some interest that we
use quantum financial models in finance theory. Indeed, some
mathematical methods on applications of quantum mechanics to
finance have been presented in [2, 3], including quantum trading
strategies, quantum hedging, and quantum version of no-arbitrage.
\\~\\
{\bf 2  The quantum model of binomial markets}

A binomial market is formed by a bank account $B = (B_0, B_1)$ and
some stock of price $S = (S_0, S_1).$ We assume that the constants
$B_0$ and $S_0$ are positive and\begin{equation}\label{e1} B_1 =
B_0 (1 + r),~~S_1 = S_0 ( 1 + R),\end{equation}where the interest
rate $r$ is a constant ($r > -1$) and the volatility rate $R$
takes two values $a$ and $b$ with\begin{equation}\label{e2} -1
\leq a< r < b.\end{equation}$A$ is uncertain and has two outcomes
and, so does $S_1.$ As said above, there are at least two models
for describing $S_1,$ one is Bernoulli's model while another is
quantum mechanical one. For reader's convenience, we include here
somewhat details on the Bernoulli's model of the binomial market
$(B, S).$

In fact, the simplest probabilistic model for $S_1$ is that
$\Omega = \{1+a, 1+b \}$ with a probability distribution $P.$
Assume that\begin{equation}\label{e3} p = P \{ S_1 = S_0 (1+b )
\}= P \{ R = b \} >0.\end{equation}It is natural to interpret the
variable $p$ as the probability of an up movement in the stock
price. The variable $1-p$ is then the probability of a down
movement. It is easy to check that there is a unique risk-neutral
measure $M$ on $\Omega$ such that$$E_M [R] = b M \{ R = b \} + a M
\{ R =  a \}=  r,$$that is,\begin{equation}\label{e4}M \{ R = b
\}= \frac{r-a}{b-a},~~M \{ R =  a \} =
\frac{b-r}{b-a}.\end{equation}Thus, the risk-neutral world of the
classical model for the binomial market has only one element $M.$

Given an option on the stock whose current price is $C.$ We
suppose that the payoff from the option is $C_b$ when the stock
price moves up to $S_0 (1+b),$ while the payoff is $C_a$ when the
stock price moves down to $S_0 (1+a).$ It is well known that by
making the portfolio riskless one has
that\begin{equation}\label{e5}C=\frac{1}{1+r} \left (
\frac{r-a}{b-a} C_b + \frac{b-r}{b-a} C_a \right
).\end{equation}This states that the value of the option today is
its expected future value with respect to the risk-neutral measure
$M$ discounted at the risk-free rate.

This result is an example of an important general principle in
option known as {\it risk-neutral valuation}. It means that for
the purposes of valuing an option (or any other derivative) we can
assume that the expected return from all traded securities is the
risk-free interest rate, and that future cash flows can be valued
by discounting their expected values at the risk-free interest
rate. We will make use of this principle in quantum domain in the
sequel.

However, a classical view, dating back to the times of J.Bernoulli
and C.Huygens, is that the expected future value with respect to
$P$ discounted at the risk-free rate$$\overline{C} = \frac{1}{1+r}
\left ( p C_b + (1-p) C_a \right )$$could be a reasonable price of
such an option (see for example [9]). It should be emphasized,
however, that this quantity depends essentially on our assumption
on the value $p.$

Therefore, it is surprising and seems counterintuitive that the
option pricing formula in equation (5) does not involve the
probabilities of the stock price moving up or down in the
classical case. It is natural to assume that, as the probability
of an upward movement in the stock price increases, the value of a
call option on the stock increases and the value of a put option
on the stock decreases. An explanation on this issue is presented
in [7, p205]. However, we would like to point out that in the
following quantum model this does not happen.

In order to propose the quantum model of the binomial market $(B,
S),$ we consider the Hilbert space ${\bf C}^2$ with its canonical
basis$$|0> = \left( \begin{array}{l}1 \\  0 \end{array}\right),
|1> = \left(
\begin{array}{l} 0 \\ 1\end{array} \right).$$Define
$$I_2 =\left( \begin{array}{ll}1 &  0 \\  0 & 1\end{array}
\right), ~\sigma_x =\left( \begin{array}{ll}0 &  1 \\  1 &
0\end{array} \right), ~\sigma_y =\left( \begin{array}{ll}0 & -i
\\  i & ~0\end{array} \right),~\sigma_z =\left(
\begin{array}{ll}1 & ~0 \\  0 & -1\end{array} \right),$$where
$\sigma_x, \sigma_y,$ and $\sigma_z$ are the well-known Pauli spin
matrices of quantum mechanics. (See [8] for details.)
Set\begin{equation}\label{e6} R = \frac{a+b}{2} I_2 + x_0 \sigma_x
+ y_0 \sigma_y + z_0 \sigma_z,\end{equation}which takes two values
$a$ and $b,$ where $x_0, y_0$ and $z_0$ are all real numbers such
that$$x^2_0 + y^2_0 + z^2_0 = \frac{(b-a)^2}{4}.$$In this case, a
quantum model for the binomial market $(B, S)$ is presented.

By the risk-neutral valuation, all individuals are indifferent to
risk in a risk-neutral world, and the return earned on the stock
must equal the risk-free interest rate. Thus, the risk-neutral
world of the quantum model $(B, S)$ consists of faithful states
$\rho$ on ${\bf C}^2$ satisfying\begin{equation}\label{e7}tr \rho
R = r.\end{equation}

Given$$\rho = \frac{1}{2} (w I_2 + x \sigma_x + y \sigma_ y + z
\sigma_z ) = \frac{1}{2} \left(
\begin{array}{ll}w + z &  x- iy \\  x+ iy & w - z\end{array}
\right),$$which takes two values
$$\lambda_1 = \frac{1}{2} \left (w - \sqrt{x^2 + y^2 + z^2} \right ),
\lambda_2 = \frac{1}{2} \left ( w + \sqrt{x^2 + y^2 + z^2}
\right ).$$Then, $\rho$ is a faithful state if and only if $t r
\rho =1$ and $\lambda_1 > 0.$ This concludes that $w = 1$ and$$x^2
+ y^2 + z^2 < 1.$$Then, by equation (7) one concludes that the
risk-neutral world of the quantum binomial model consists of
states of the form
\begin{equation}\label{e8}\rho = \frac{1}{2}
(I_2 + x \sigma_x + y \sigma_ y + z \sigma_z ),\end{equation}where
all $(x,y,z)$ satisfy\begin{equation}\label{e9} \left \{
\begin{array}{l} x^2 + y^2 + z^2 < 1,\\x_0 x + y_0 y + z_0 z = r
- \frac{a+b}{2},
\end{array} \right. \end{equation}which is a disk of radius
$\sqrt{1- \frac{(2 r - a-b)^2}{(b-a)^2}}$ in the unit ball of
${\bf R}^3.$ Moreover, the quantum binomial model is
arbitrage-free if and only if $ -1 \leq a < r < b.$

Equations (8) and (9) characterize the risk-neutral world of the
quantum binomial market. In contrast to the classical binomial
market whose risk-neutral world consists of only one element $M$
in (4), this quantum risk-neutral world has infinite elements. It
is open and its size depends on the risk-free rate $r,$ which
attains the maximum at $r = \frac{a+b}{2}.$

By (7) under any risk-neutral state $\rho$ the probabilities of
that $R$ takes value $b$ and $a$ are $\frac{r-a}{b-a}$ and
$\frac{b-r}{b-a}$ respectively, the current price of the option on
the stock is thus $C$ in (5) by the risk-neutral valuation.
Therefore, we obtain the same result by using the quantum model
without assuming the probabilities of the stock price moving up or
down as in the classical case.

A more concrete example is the European call option in the quantum
binomial market with the exercise price $K.$ Its payoff is of the
form$$H = (S_1 - K)^+,$$which takes two values$$h_a = \max(0, S_0
(1 + a) - K),~~h_b = \max(0, S_0 (1 + b) - K).$$Thus, the option
value $C$ today is
\begin{equation}\label{e10}C = \frac{1}{1+ r} tr [ \rho H] = \frac{1}{1+ r}
\left ( \frac{b-r}{b-a} h_a + \frac{r-a}{b-a} h_b \right
)\end{equation}for all states $\rho$ in the risk-neutral world.
\\~\\
{\bf 3  Cox-Ross-Rubinstein binomial option pricing formula via
quantum mechanics}

In the early 1970s F.Black, M.Scholes, and R.C.Merton made a major
breakthrough in the pricing of stock options, see [1] and [6].
This involved the development of what has become known as the
Black-Scholes model, in which the famous ``Black-Scholes Option
Pricing Formula" was derived. In 1979 J.C.Cox, S.A.Ross, and
M.Rubinstein [4] presented a widely used numerical procedure for
the Black-Scholes option pricing formula, by dividing the life of
the option into a large number of small time intervals. They
argued that the Black-Scholes model is the limitation of models of
a much large numbers of small binomial markets, in which the
famous Cox-Ross-Rubinstein binomial option pricing formula was
found. From the physical view of point we find that J.C.Cox,
S.A.Ross, and M.Rubinstein [4] used the classical model of
multi-period binomial markets for obtaining their formula. In the
following we will use a quantum model of multi-period binomial
markets to re-deduce the Cox-Ross-Rubinstein binomial option
pricing formula.

Consider Maxwell-Boltzmann statistics of the system of $N$
distinguishable particles of two-level energies $a$ and $b.$ The
mathematical model is then as following: Let ${\bf H}_n = ( {\bf
C}^2 )^{\otimes n}$ and write$$ |\varepsilon_1 ... \varepsilon_n>
= |\varepsilon_1
> \otimes ... \otimes |\varepsilon_n>,~~\varepsilon_1 ,...,
\varepsilon_n = 0,1.$$ Then, $\{ |\varepsilon_1 ... \varepsilon_n>
: \varepsilon_1 ,..., \varepsilon_n = 0,1 \}$ is the canonical
basis of ${\bf H}_n.$ Given $-1 \leq a < r < b,$ we define a
$N$-period quantum binomial market $(B,S)$ with $B=(B_0,
B_1,...,B_N)$ and $S=(S_0, S_1,...,S_N)$ as following:
\begin{equation}\label{e11}B_n = B_0 (1 + r)^n,~~S_n = S_0
\bigotimes^n_{j=1} (1 + R_j) \otimes I_{ N -
n},~~n=1,...,N,\end{equation} where the constants $B_0$ and $S_0$
are positive, $I_{N-n}$ is the identity on ${\bf H}_{N-n}$ and,
$$R_j = \frac{a+b}{2} I_2 + x_{0j} \sigma_x + y_{0j} \sigma_y + z_{0j}
\sigma_z,$$where$$x^2_{0j} + y^2_{0j} + z^2_{0j} =
\frac{(b-a)^2}{4}$$for all $j=1,...,N.$

Consider European call options in the $N$-period quantum binomial
market $(B, S).$ Its payoff is $$H_N = (S_N - K)^+,$$where $K$ is
the exercise price of the European call option. There are $N$
orthonormal bases $\{ (u_j, v_j ),j=1,...,N \}$ in ${\bf C}^2$
such that$$\begin{array}{lcl}S_N & =& S_0 \bigotimes^N_{j=1} (1 +
R_j )\\~\\&=& S_0 \bigotimes^N_{j=1} \left [ (1 + b) |u_j >< u_j |
+ (1 + a) |v_j >< v_j | \right ]\\~\\& = & S_0 \sum^N_{n=0} (1 +
b)^n (1 + a)^{N-n} \left [ \sum_{| \sigma | = n}
\bigotimes^N_{j=1} | w_{j \sigma}
>< w_{j \sigma} | \right ]\end{array}$$where all $\sigma$ are subsets of
$\{ 1,..., N \}, w_{j \sigma} = u_j$ for $j \in \sigma$ or $w_{j
\sigma} = v_j$ otherwise. Hence$$(S_N - K)^+ = \sum^N_{n=0} \left
[ S_0 (1 + b)^n (1 + a)^{N-n} - K \right ]^+ \left [ \sum_{|
\sigma | = n} \bigotimes^N_{j=1} | w_{j \sigma}
>< w_{j \sigma} | \right ].$$

Since the $N$ distinguishable particles are all free, all states
of the form\begin{equation}\label{e12}\bigotimes^N_{j=1} \rho_j =
\frac{1}{2^N} \bigotimes^N_{j=1} \left ( I_2 + x_j \sigma_x + y_j
\sigma_ y + z_j \sigma_z \right )\end{equation}are faithful
risk-neutral states of the $N$-period quantum binomial market $(B,
S),$ where $(x_j, y_j, z_j)$ satisfies$$\left\{ \begin{array}{l}
x^2_j + y^2_j + z^2_j < 1,\\x_{0j} x_j + y_{0j} y_j + z_{0j} z_j =
r - \frac{a+b}{2},
\end{array} \right.$$for every $j=1,...,N.$ Moreover, by using
the Maxwell-Boltzmann statistics, one has
that\begin{equation}\label{e13}tr \left [ \left (
\bigotimes^N_{j=1} \rho_j \right ) \sum_{| \sigma | = n}
\bigotimes^N_{j=1} | w_{j \sigma}
>< w_{j \sigma} | \right ] = \frac{N !}{n ! (N - n) !} q^n (1-q)^{N -
n}\end{equation}for $n=0,1,..., N,$ where $q = \frac{r - a}{b-a}.$

Therefore, by the principle of risk-neutral valuation, the price
$C^N_0$ at time $0$ of the European call option $(S_N - K)^+$ is
given by$$\begin{array}{lcl}C^N_0 & = & (1 + r)^{-N} tr \left [
\left ( \bigotimes^N_{j=1} \rho_j \right ) (S_N - K)^+ \right
]\\~\\& = & (1 + r)^{-N} \sum^N_{n=0} \frac{N !}{n ! (N - n) !}
q^n (1-q)^{N - n} \left [ S_0 (1 + b)^n (1 + a)^{N-n} - K \right
]^+\\~\\& = & S_0 \sum^N_{n = \tau} \frac{N !}{n ! (N - n) !} q^n
(1-q)^{N - n} \frac{(1 + b)^n (1 + a)^{N-n}}{(1 + r)^N}\\~\\&~~& -
K (1 + r)^{-N} \sum^N_{n = \tau} \frac{N !}{n ! (N - n) !} q^n
(1-q)^{N - n},\end{array}$$where $\tau$ is the first integer $n$
for which$$S_0 (1+ b)^n (1 + a)^{N-n} > K.$$

Now observe that using $q = \frac{r - a}{b-a}$ and$$q^{\prime} = q
\frac{1 + b}{1 + r}$$we obtain $0 < q^{\prime} < 1$ so that we can
finally write the fair price for the European call option in this
multi-period quantum binomial pricing model
as\begin{equation}\label{e14}C^N_0 = S_0 \Psi (\tau ; N,
q^{\prime} ) - K (1 + r)^{-N} \Psi (\tau ; N,
q)\end{equation}where $\Psi$ is the complementary binomial
distribution function, that is,$$\Psi (m; n, p) = \sum^n_{j=m}
\frac{n !}{j ! (n-j) !} p^j ( 1-p)^{n-j}.$$This is just the {\it
Cox-Ross-Rubinstein binomial option pricing formula}.
\\~\\
{\bf 4  Conclusion}

The binomial markets are hypothesis and very imprecise models. It
was J.C.Cox, S.A.Ross, and M.Rubinstein [4] who concluded that a
realistic model can be regarded as a limitation of a much large
numbers of small binomial markets. Their approach to the binomial
markets is by using classical probability methods. However, in
nature there are some binomial problems which cannot be described
by classical random variables. For example, photon's polarization
must be described by quantum mechanics. This indicates that we may
interrupt the binomial markets as some quantum models. We have
shown that the risk-neutral world of the quantum binomial markets
exhibits an intriguing structure as a disk in the unit ball of
${\bf R}^3,$ whose radius is a function of the risk-free interest
rate with two thresholds which prevent arbitrage opportunities
from this quantum market. Moreover, we re-deduce the
Cox-Ross-Rubinstein binomial option pricing formula by considering
Maxwell-Boltzmann statistics of the system of $N$ distinguishable
particles as a model of the multi-period binomial markets.

We would like to mention that besides the Maxwell-Boltzmann
statistics, there are Bose-Einstein statistics of the system of
$N$ identical particles in quantum mechanics. When consider a
many-particle system satisfying Bose-Einstein statistics as a
model of the multi-period binomial markets, we will give another
binomial option pricing formula as following.

Indeed, set$$S_N = S_0 (1 + R)^{ \hat{\otimes} N},~R  =
\frac{a+b}{2} I_2 + x_0 \sigma_x + y_0 \sigma_y + z_0
\sigma_z,$$where $x^2_0 + y^2_0 + z^2_0 = \frac{(b-a)^2}{4}.$
Given any risk-neutral state $\rho$ of the binomial market, from
the quantum view of point it is easy to see that
$\rho^{\hat{\otimes} N}$ is a risk-neutral state of the identical
particle model of the multi-period binomial markets. Then, by
using the risk-neutral valuation we conclude from the
Bose-Einstein statistics that the price $C^N$ today of the
European call option $(S_N - K)^+$ in the present model
is$$\begin{array}{lcl}C^N & = & (1 + r)^{-N} tr \left [
 \rho^{\hat{\otimes} N} (S_N - K)^+ \right ]\\~\\& = &
\frac{1}{(1 + r)^N } \sum^N_{n=0} \left ( \frac{q^n (1-q)^{N-n}
}{\sum^N_{k=0} q^k (1-q)^{N-k} } \right ) \left [ S_0 (1 + b)^n (1
+ a)^{N-n} - K \right ]^+.\end{array}$$We hope that this binomial
option pricing formula will be found to be useful in finance,
since the Bose-Einstein statistics plays a crucial role in quantum
mechanics as the same as the Maxwell-Boltzmann statistics.
\\~\\
{\bf Acknowledgments.} This paper was revised during the author
visited Academy of Mathematics and Systems Science, Chinese
Academy of Sciences. The author is very grateful to Professor
Shouyang Wang for his comments and advice.

\begin{center}{\bf References}\end{center}
\begin{description}
\item[1] F.Black, M.Scholes, The pricing of options and corporate
liabilities, {\it Journal of Political Economy,} 1973, {\bf 81}:
637-659 \item[2] Zeqian Chen, The meaning of quantum finance (in
Chinese), {\it Acta Mathematica Scientia,} 2003, {\bf 23}A(1):
115-128 \item[3] Zeqian Chen, Quantum finance: The finite
dimensional case, www.arxiv.org/quant-ph/0112158 \item[4] J.C.Cox,
S.A.Ross, M.Rubinstein, Option pricing: a simplified approach,
{\it Journal of Finance Economics,} 1979, {\bf 7}(3): 229-263
\item[5] P.A.M.Dirac, {\it The Principles of Quantum Mechanics,}
Oxford University Press, Oxford, 1958 \item[6] R.C.Merton, {\it
Continuous-time Finance,} Basil Black-Well, Cambridge, MA, 1990
\item[7] J.C.Hull, {\it Options, Futures and Other Derivatives,}
4th Edition, Prentice-Hall, Inc., Prentice, 2000 \item[8]
K.P.Parthasarathy, {\it An Introduction to Quantum Stochastic
Calculus,} Birkh\"{a}user Verlag, Basel, 1992 \item[9]
A.N.Shiryaev, {\it Essentials of Stochastic Finance $($Facts,
Models, Theory$)$,} World Scientific, Singapore, 1999
\end{description}
\end{document}